\documentstyle[mprocl,psfig]{article}
\def\be{\begin{equation}}
\def\ee{\end{equation}}
\def\ba{\begin{eqnarray}}
\def\ea{\end{eqnarray}}
\newcommand{\bec}{\begin{center}}
\newcommand{\eec}{\end{center}}

\begin{document}

\title{BOSON STARS: EARLY HISTORY AND RECENT PROSPECTS\footnote{\ \ Report
of parallel session chair in:
Proc. 8th M. Grossmann Meeting, T. Piran (ed.), World Scientific,
Singapore 1998, to be published}}

\author{
ECKEHARD W. MIELKE\footnote{\ \ E-mail address: {\tt ekke@xanum.uam.mx}}}

\address{\footnotesize
Departamento de F\'{\i}sica,\\
Universidad Aut\'onoma Metropolitana--Iztapalapa,\\
Apartado Postal 55-534, C.P. 09340, M\'exico, D.F., MEXICO}

\author{
FRANZ E. SCHUNCK\footnote{\ \ E-mail address: {\tt fs@astr.cpes.susx.ac.uk}}}

\address{\footnotesize
Astronomy Centre, University of Sussex,\\
Falmer, Brighton BN1 9QJ, United Kingdom}

\maketitle\abstracts{
Boson stars  are
descendants of the so--called {\em geons} of Wheeler, except that 
they are built from  scalar particles 
instead of electromagnetic fields.  
If scalar fields exist in nature,  such localized  configurations 
kept together by their self-generated gravitational field 
can  form within Einstein's general relativity.  
In the case of {\em complex} scalar fields, an
{\em absolutely stable}  branch of such non-topological solitons 
with conserved particle number exists.\\
Our present surge stems 
from the speculative possibility that these compact objects could provide a
considerable fraction of the
non-baryonic part of dark matter. In any case, they may serve as a 
convenient  ``laboratory"
for studying numerically rapidly {\em rotating} bodies in general relativity
and the generation of gravitational waves.}

\section{Introduction}
If scalar fields exist in nature, soliton-type configurations kept together
by their self-generated gravitational field 
can form absolutely stable {\em boson stars} (BS), resembling neutron stars. 
They are descendants of the so--called {\em geons} of Wheeler  \cite{W55}.

We will review the history of these {\em hypothetical} stars, starting  1968
with the work of Kaup as well as that of Ruffini and Bonazzola.
In building macroscopic boson stars,  a nonlinear Higgs type 
potential was later  considered
as an additional  repulsive interaction. 
Thereby the Kaup limit  for boson stars
can  even {\em exceed}  the limiting mass of 3.23 $M_\odot$
for  neutron stars. 

Moreover, in the spherically
symmetric case, we have shown via catastrophe theory \cite{kus} that these
{\em boson stars} have a {\em stable branch} with a wide range of 
masses and radii.

Recently, we construct \cite{SM96,MS96} for the first time the 
corresponding  localized {\em rotating} configurations 
via numerical integration of the coupled 
Einstein--Klein--Gordon equations. Due to {\em gravito--magnetic} effect,
the ratio of conserved angular momentum and particle number turns out 
to be an integer $a$, the {\em azimuthal quantum number} of our 
soliton--type stars. The resulting axisymmetric metric, the energy density 
and the Tolman mass are {\em completely regular}. Moreover, we analyze 
the {\em differential rotation} and {\em stability} of such fully 
relativistic configurations.

The present surge stems from the possibility that
these localized objects could provide a considerable fraction of the 
non-baryonic part of dark matter.

\subsection{Geons in general relativity}
Transfering the ideas 
of Mach and Einstein to the microcosmos, the curving up of the 
background metric should be self-consistently produced by the stress-energy
content $T_{\mu\nu}$   of matter  
via the Einstein equations with cosmological term. In some 
geometrodynamical models \cite{M77,M78}, extended particles 
owning internal symmetries were
classically described by objects which closely resemble  
{\em geons}
or wormholes. 
The {\em geon}, i.e.~a {\em g}ravitational {\em e}lectromagnetic entity, was originally devised
by Wheeler \cite{W55} to be a self-consistent, nonsingular solution of 
the otherwise source-free Einstein--Maxwell equations having persistent 
large-scale features.
It realizes to some extent 
the proposal of Einstein and Rosen in  
their 1935 paper \cite{ER35}:

``Is an atomistic theory of matter and electricity conceivable which, while
excluding singularities in the field, makes use of no other fields than those
of the gravitational field $(g_{\mu\nu})$ and  those of the 
electromagnetic field in the
sense of Maxwell  (vector potentials $\phi_\mu$)?"

The Lagrangian density of gravitational coupled Maxwell 
field $A:= A_\mu\, dx^\mu$ reads
\be
{\cal L}_{\rm geon} = \frac{1}{2\kappa} \sqrt{\mid g\mid}  R  -
\frac{1}{4}  \sqrt{\mid g\mid} F_{\mu \nu} \,F^{\mu \nu}\, ,  
\label{geonlag}
\ee
where $\kappa = 8\pi G$ is the gravitational constant in natural units,
$g$ the determinant of the metric $g_{\mu \nu }$,
$R:= g^{\mu \nu } R_{\mu \nu }
= g^{\mu \nu } \Bigl (
\partial_\nu \Gamma_{\mu \sigma }{}^\sigma
- \partial_\sigma \Gamma_{\mu \nu }{}^\sigma +
\Gamma_{\mu \sigma }{}^\alpha \Gamma_{\alpha \nu }{}^\sigma
- \Gamma_{\mu \nu }{}^\alpha \Gamma_{\alpha \sigma }{}^\sigma \Bigr )$
the curvature scalar with  Tolman's sign convention \cite{T34}, and 
$F=dA=(1/2) F_{\mu\nu}\, dx^\mu\wedge dx^\nu$. Greek indices
$\mu ,\nu, \cdots$ are running from 0 to 3.

Such a geon provides a well-defined model for a classical body in 
general relativity exhibiting ``mass without mass".
If spherically symmetric geons would stay  stable, 
the possibility would arise to derive the equations of 
motions \cite{W61} for the center of gravity solely from Einstein's
field equations  without the need to introduce
field singularities. In a sense this approach also achieves some of the 
 goals of the
so-called unitary field theory \cite{F49,FLR51}.

Geons, as we are using the term, are {\em gravitational solitons}, which are 
held together by self-generated gravitational forces and are composed of 
localized fundamental classical fields. The coupling of gravity to 
neutrino fields
has already been considered by Brill and Wheeler \cite{BW57}. It 
lays the appropiate groundwork for an extension 
to nonlinear {\em spinor  geons} satisfying the combined 
Einstein--Dirac equations. In previous papers \cite{DM79,MS81}
however,  algebraic complications resulting from the spinor 
structure as well as from the internal symmetry are avoided  
by considering, instead, 
interacting   {\em scalar fields} coupled to gravity. In order to maintain a 
similar dynamics, a scalar  self-interaction $U(\Phi)$ is assumed 
 which can be formally obtained by 
"squaring" the fundamental nonlinear spinor equation. 
Klein--Gordon geons
have been previously constructed by Kaup \cite{K68}. However, 
the additional nonlinearity 
of the scalar fields turns out to be an important new ingredient.

\subsection{Do scalar fields exist in nature?}
The physical nature  
of the  spin--0--particles out of which the boson star (BS) is presumed to 
consist, is still an open issue.
Until now, no fundamental elementary scalar particle has been found in 
accelerator experiments, 
which could serve as the main constituent of the boson star.
 In the theory of Glashow, Weinberg, and Salam, 
a Higgs boson--dublett $(\Phi^+, \Phi^0)$ and its anti-dublett
$(\Phi^-, \bar \Phi^0)$ are necessary ingredients in order
to generate masses for the $W^{\pm }$ and $Z^0$ gauge vector bosons 
\cite{[5]}. After symmetry breaking, only one scalar particle, the
Higgs particle 
$H:=(\Phi^0 + \bar \Phi^0)/\sqrt{2}$, remains free and occurs as a state
in a constant scalar field background \cite{perkins}. Nowadays, as
it is indicated by 
the rather heavy top quark \cite{abe} of 176  GeV/$c^2$, one 
expects the mass of the Higgs particle to be close to  1000 GeV/$c^2$.
However, above 1.2 TeV/$c^2$ the self--interaction $U(\Phi)$ of the 
Higgs field is so large that the perturbative approach of the 
standard model becomes unreliable. Therefore a conformal extension of 
the standard model with gravity included may be 
necessary, see \cite{Paw,hehl}.
High--energy experiments at the LHC at Cern will reveal 
if these Higgs particles really exist in nature. 

As free  particles, the Higgs boson
is unstable with respect to the decays  $H \rightarrow W^+ + W^-$ and 
$H \rightarrow Z^0+ Z^0$. In a compact
object like the boson star, these decay channels are expected to be in 
equilibrium with the inverse process $Z^0+ Z^0 \rightarrow H$,
for instance. This is presumably in full analogy with the neutron star
\cite{weigel} or quark star \cite{weber,glenn},
where one finds an equilibrium of $\beta $-- and inverse $\beta $--decay 
of the neutrons or quarks and thus stability of the macroscopic star with 
respect to radioactive decay.
Nishimura and Yamaguchi \cite{[6]}  constructed a
neutron star using an equation of state of an isotropic fluid built 
from Higgs bosons.

\section{Boson stars}
In a perspective paper Kaup \cite{K68} has studied for the first time  
the full generally relativistic coupling of linear Klein--Gordon fields 
to gravity in a localized configuration.
It is already realized that  
no Schwarzschild type 
event horizon occurs in such  numerical solutions. Moreover, the problem of 
the stability of 
the resulting scalar geons with respect to radial perturbations is treated. 
It is shown that such objects are resistant to gravitational collapse (related 
works include Refs.~\cite{D63,FM68,TWS75}). These considerations are 
on a semiclassical level, since the Klein--Gordon field is treated as a 
classical field. However, using a Hartree--Fock approximation for the second
quantized two--body problem, Ruffini and Bonazzola \cite{RB69} showed  that 
the same coupled 
Einstein--Klein--Gordon equations apply.  Exact but singular solutions of 
the coupled Maxwell--Einstein--Klein--Gordon equation have been constructed 
before by Das \cite{D63}.

The Lagrangian density of gravitationally coupled complex scalar 
field $\Phi$ reads
\be
{\cal L}_{\rm BS} = \frac {\sqrt{\mid g\mid }}{2\kappa } \left \{ R
 + \kappa 
   \left [ g^{\mu \nu} (\partial_\mu \Phi^*) (\partial_\nu \Phi )
             - U(\mid \Phi \mid^2) \right ] \right \} \, .  
             \label{lagrange}
\ee
Using the principle of variation, one finds the coupled
Einstein--Klein--Gordon equations
\ba
 G_{\mu \nu }:=  R_{\mu \nu } - \frac{1}{2} g_{\mu \nu } R
             & = & -\kappa T_{\mu \nu } (\Phi ) \; , \label{phi152} \\
      \left (\Box + \frac {dU}{d\mid \Phi \mid^2} \right ) \Phi
             & = & 0 \; , \label{phi153}
\ea
where
\be
T_{\mu \nu }(\Phi )
 = {1\over2} [ (\partial_\mu \Phi^\ast )(\partial_\nu \Phi )
  + (\partial_\mu \Phi )(\partial_\nu \Phi^\ast ) ]
  - g_{\mu \nu } {\cal L} (\Phi )/\sqrt {\mid g\mid }
\ee
is the {\em stress--ener\-gy ten\-sor} and
$
\Box := \left (1/\sqrt {\mid g\mid }\right )\,
 \partial_\mu  \left (\sqrt{\mid g\mid } g^{\mu \nu }
 \partial_\nu \right )
$
the generally covariant d'Alembertian.

The stationarity ansatz
\be
\Phi (r,t)=P(r) e^{-i\omega t} 
\ee
 describes a spherically symmetric bound 
state of the scalar field with frequency $\omega $.

In the case of spherical symmetry, the line-element 
reads
\be
ds^2 = e^{\nu (r)} dt^2 - e^{\lambda (r)} \biggl [ dr^2 + r^2 \Bigl (
 d\theta^2 + \sin^2\theta d\varphi^2 \Bigr ) \biggr ] \; ,
\ee
in which the functions $\nu =\nu (r)$ and 
$\lambda =\lambda (r)$ depend on the Schwarzschild type radial
coordinate $r$.

In the years following the geons of Wheeler \cite{W55},
some efforts were also made in order to find a (semi--) classical 
model describing elementary particles. 
In 1968, Kaup \cite{K68} presented the notion of the  
`Klein--Gordon geon', which nowadays has been christened
{\em mini--boson star}. It can be regarded  as a 
{\em macroscopic quantum state}.

As in 
the case of a prescribed Schwarzschild background \cite{DM79}, the spacetime  
curvature affects the resulting Schr\"o\-din\-ger equation for the
radical function $P(r)$ essentially via an external gravitational potential.
Indeed Feinblum and McKinley \cite{FM68} found eigensolutions with
nodes corresponding to the principal 
quantum number $n$ of the H--atom. They also realized that localized solutions 
fall off asymptotically as 
$P(r) \sim (1/r)\exp\left(-\sqrt{m^2 -E^2}\, r\right)$ in a 
Schwarzschild-type asymptotic background.

The energy--momentum tensor becomes diagonal,
i.e.~$T_\mu{} ^\nu(\Phi) = {\rm diag} \; (\rho , -p_r,$
$-p_\bot, -p_\bot )$ with
\ba
\rho &=& \frac{1}{2} (\omega^2 P^2 e^{-\nu} +P'^2 e^{-\lambda} +U )\; ,
\nonumber  \\
p_r &=&  \rho -  U \; , \;  \nonumber  \\ 
p_\bot &=&  p_r -  P'^2 e^{-\lambda } \; .   
 \ea
This  form  is familiar from  fluids,
except that the radial and tangential pressure generated by the
scalar field are in general different, i.e.~$p_r \neq p_\bot $, 
due to the different sign of  
$(P')^2$ in these expressions. 

In general, the resulting system of three coupled nonlinear equations for the
radial parts of the scalar and the (strong) gravitational tensor field has
to be solved numerically. In order to specify the starting values for the 
ensuing numerical analysis, asymptotic solutions at the
origin and at spatial infinity are instrumental. 

That the stress--energy tensor of  a 
BS, unlike a classical fluid,  is in general {\em anisotropic} 
has already been noticed by Kaup \cite{K68}. 
In contrast to neutron stars \cite{HT65,[Zel]}, where the ideal fluid
approximation demands an
isotropic symmetry for the pressure, for spherically symmetric 
boson stars there are different stresses $p_r$ and $p_\bot$  in radial or 
tangential directions, respectively. 
Ruffini and Bonazzola \cite{RB69} introduced the notion of 
{\it fractional anisotropy}\ $a_f :=(p_r - p_\bot )/p_r =
P'^2 e^{-\lambda}/(\rho -U)$ which depends essentially 
on the self-interaction; cf.~Ref.~\cite{Ge88}.

So the perfect fluid approximation is totally 
inadequate for boson stars. Actually, Ruffini and
Bonazzola \cite{RB69,[4]} used the  formalism of  second quantization 
for the complex Klein--Gordon field and observed an important feature: If all
scalar particles are within the {\em same ground state}
$|\Phi > =(N,n,l,a)=(N,0,0,0)$, which 
is possible
because of Bose--Einstein statistics, then the semi--classical
Klein--Gordon equation of Kaup is recovered in the 
Hartree--Fock approximation.   In contrast to the Newtonian 
approximation, the full relativistic 
treatment avoids an unlimited increase of the 
particle number and negative energies, but induces 
critical masses and particle numbers with a global maximum.

There exists a decisive difference between self--gravitating 
objects made of fer\-mions or bosons:
For a many fermion system the Pauli exclusion principle 
forces the typical fermion 
into a state with very high quantum number, whereas many bosons can 
coexist all in the same ground state (Bose--Einstein condensation). 
This also reflects itself 
in the critical number of stable configurations:
\begin{itemize}
\item{}
$N_{\rm crit} \simeq (M_{\rm Pl}/m)^3$ for fermions 
\item{}
$N_{\rm crit} \simeq (M_{\rm Pl}/m)^2$ for massive bosons without 
self--interaction. 
\end{itemize}  

Cold mixed boson--fermion stars have been studied by 
Henrique et al.~\cite{HLM89} and 
Jetzer \cite{Je90}.
\subsection{Gravitational atoms as boson stars}

In a nut--shell, a boson star is a stationary solution of a 
(non-linear) Klein--Gordon equation  in its own 
gravitational field; cf.~\cite{Mi78,Mi80a}.
We treat this problem in a {\em semi-classical} manner,
because effects of the quantized gravitational field are neglected. 
Therefore, a (Newtonian) boson star is also 
 called a {\it gravitational atom} \cite{FG89}. Since a 
 free Klein--Gordon equation for a complex scalar field is a 
 {\em relativistic generalization} of the 
 {\em Schr\"odinger equation}, we consider for the {\it ground state} 
 a generalization of the wave function
 \ba
 \vert N, n, l,a>: \qquad \Phi&=& R_a^n(r)\, 
 Y^{|a|}_l(\theta, \, \varphi)e^{-i (E_n/\hbar)t} \nonumber\\
 &=&{1\over\sqrt{ 4\pi}} R_a^n(r)
 P^{|a|}_l(\cos\theta)\, e^{ia\varphi}\, 
e^{-i (E_n/\hbar)t} \label{Hatom}
\ea 
of the hydrogen atom. Here $R_a^n(r)$ is the radial distribution,
$Y^{|a|}_l(\theta, \, \varphi)$ the spherical harmonics,  
$P^{|a|}_l(\cos\theta)$ are the 
normalized Legendre polynomials, and $|a|\leq l$ are  the quantum numbers
of {\em azimuthal}  and {\em angular} momentum.
 
Thus `gravitational atoms' represent {\em coherent} quantum states, which
nevertheless can have macroscopic size and large masses. 
The gravitational field is self-generated 
via the energy--momentum tensor, but remains 
completely classical, whereas the complex
scalar fields are treated to some extent as Schr\"odinger 
wave functions, which in quantum field theory are 
referred to as semi-classical.

Motivated by Heisenberg's non-linear spinor equation \cite{H66,Mi81}
additional self--inter\-acting terms   describing the interaction 
between the bosonic particles in a ``geon'' type configuration were first 
considered by  Mielke and Scherzer \cite{MS81}, where also solutions
with nodes, i.e.~``principal quantum number'' $n>1$ and non-vanishing
angular momentum  $l \neq 0$ 
for a t'Hooft type monopole ansatz $\Phi^I \sim R(r)\, P^{|I|}_l(\cos\theta)$ 
were found. An analytical solution ~\cite{BMHH87} of
the coupled Einstein-scalar-field system  also  exists.
Recently Rosen \cite{rosen} reviewed his old idea of an elementary 
particle built
out of scalar fields  within the framework of the Klein--Gordon geon  or 
the mini--boson star.

Further analysis is needed in order to understand these highly interesting
instances of a possible fine structure in the energy levels of 
gravitational atoms. In
view of these rich and prospective structures, are
{\em quantum geons} \cite{W61}  
capable of internal excitations?

In building macroscopic boson
stars, Colpi et al.~\cite{colpi} used a Higgs--type self-interaction in order  
to accommodate a repulsive
force besides  gravity. This
repulsion between the constituents is instrumental to
blow up the boson star so that much more particles will have
room in the confined region. Thus the maximal mass of a BS 
can reach or even extend the limiting mass of 3.23 $M_\odot$
for  neutron stars \cite{fried,cook} with {\em realistic} equations of
state $p=p(\rho)$ for which
the (phase) velocity of sound is $v_s=\sqrt{dp/d\rho}\leq c$. However,
this fact depends on the strength of the self--interaction.
The exciting possibilty of having cold stars with very large $n$ may add 
another thread to the question of black hole formation.

The work of Friedberg et al.~\cite{[B2]} renewed the interest
in the study of  boson stars.
They investigated  the Newtonian limit, analyzed in more detail the 
solutions with higher nodes of Feinblum et al.~\cite{FM68} and 
Mielke and Scherzer \cite{MS81}, and
in a preliminary form, stability questions.
Several  surveys \cite{jetzer,lee,strau}
summarize the present status of the non-rotating case.

\subsection{Critical masses of boson stars}

The Noether theorem associates with each 
symmetry a {\em locally conserved current} $\partial_\mu j^\mu =0$ and 
``charge''.
The first ``constant of motion'' of our coupled system of equations 
is given by the invariance of the Lagrangian density under a 
global phase transformation
$\Phi \rightarrow \Phi e^{-i\vartheta }$ of the complex scalar field. 
{}From the 
associated Noether current $j^\mu$ arises  the
{\em particle number}:
\be
N := \int j^0  d^3x
\; , \qquad\qquad
j^\mu = \frac {i}{2} \sqrt{\mid g\mid }\; g^{\mu \nu }
 [\Phi^\ast \partial_\nu \Phi -\Phi \partial_\nu \Phi^\ast ] \; . 
\label{teilchen} 
\ee
For the {\em total gravitational mass} of localized solutions we use 
Tolman's expression \cite{T30,G}:
\be
M  =  \int (2T_0^{\; 0}-T_\mu^{\; \mu }) 
         \sqrt{\mid g\mid} \; d^3x \; . \label{Tolm}
\ee

Since boson stars are {\em macroscopic quantum states}, they are 
prevented from complete
gravitational collapse by the Heisenberg uncertainty principle.

This provides us also with  crude mass estimates: For a boson to be confined 
within the star of  radius $R_0$, the Compton wavelength has to satisfy
$\lambda_\Phi= (2\pi\hbar/mc) \leq 2R_0$. On the other hand,
the star's radius    
should be of the order of   the last stable Kepler orbit 
$3R_{\rm S}$ around a black hole of Schwarzschild radius  
$R_{\rm S}:= 2GM/c^2$ in order to avoid an instability 
against complete gravitational  collapse.

For a  {\em mini--boson star} of effective  
radius  $R_0 \cong (\pi/2)^2 R_{\rm S}$ close to its Schwarz\-schild radius 
one obtains the estimate
\be 
M_{\rm crit} \cong (2/\pi)M_{\rm Pl}^2/m \geq  0.633\, M_{\rm Pl}^2/m\, ,
\ee
cf.~Ref.~\cite{jetzer}, which provides a rather good upper bound on the 
so-called {\em Kaup limit}. 
The correct value in the second expression 
was found only numerically as a limit of  the 
maximal mass of a {\em stable} mini--boson star. Here $M_{\rm Pl}:=\sqrt{\hbar c /G}$ 
is the Planck mass 
and $m$ the mass of a bosonic particle. For a mass of $m=30$ GeV/c$^2$, 
one can estimate 
the total mass of this mini--boson star to be $M\simeq 10^{10}$ kg and its 
radius $R_0\simeq 10^{-17}$ m. This amounts to a 
density 10$^{48}$ times that of a neutron star.

This result was later extended by Colpi et al.~\cite{colpi} for the bosonic 
potential  
\be
U(\vert\Phi\vert) =m^2\vert \Phi\vert^2 +
(\lambda/2)\vert\Phi\vert^4 \,.
\ee
with  an additional quartic self--interaction.     
Since $ \vert \Phi\vert\sim  M_{\rm Pl}/\sqrt{8\pi}$ inside the boson star, one finds 
the energy density
\be
\rho \simeq m^2 M_{\rm Pl}^2\left(1 + \Lambda/8\right) \, , 
\quad {\rm where} \quad  
\Lambda:=  {\lambda \over 4\pi}  {M_{\rm Pl}^2 \over m^2}\, .
\ee
This corresponds to a  star formed from  non--interacting bosons with 
rescaled mass 
$m\rightarrow m/\sqrt{1 +  \Lambda/8} $.
Consequently, the maximal mass of a  {\em stable} BS scales 
with the coupling constant $\Lambda$ 
approximately as 
\be 
M_{\rm crit} \simeq {2\over\pi}\sqrt{1 +  \Lambda/8} 
{M_{\rm Pl}^2 \over m} \quad \rightarrow \quad
 {1\over{\pi\sqrt{2}}}\sqrt{\Lambda} {M_{\rm Pl}^2 \over m} \qquad {\rm for}\quad 
 \Lambda \rightarrow \infty
\, ,
\ee
cf.~Fig.~2 from Colpi et al.~\cite{colpi}.

For $m \simeq 1$ GeV/c$^2$ of the order of the proton mass
and $\Lambda \simeq 1$, this is in the range of the 
Chandrasekhar limiting mass $M_{\rm Ch}:=M_{\rm Pl}^3/m^2\simeq 1.5 
M_\odot$,  
where $M_\odot$ denotes the mass of the sun. 

$$\vbox{\offinterlineskip
\hrule
\halign{&\vrule#&\strut\quad\hfil#\quad\hfil\cr
height2pt&\omit&&\omit&&\omit&\cr
& {\bf Compact} && {\bf Critical mass} && {\bf Particle Number } &\cr
&{\bf Object}  && $M_{\rm crit}$ && $N_{\rm crit}$ &\cr
height2pt&\omit&&\omit&&\omit&\cr
\noalign{\hrule}
height2pt&\omit&&\omit&&\omit&\cr
& Fermion Star:  &&  $M_{\rm Ch}:=M_{\rm Pl}^3/m^2$  &&   $\sim(M_{\rm Pl}/m)^3$ &\cr 
& Mini--BS:  &&  $M_{\rm Kaup}=0.633\, M_{\rm Pl}^2/m $  &&   $0.653 \,(M_{\rm Pl}/m)^2$ &\cr 
& Boson Star:  &&  $(1/\pi\sqrt{8\pi})\sqrt{\lambda} M_{\rm Pl}^3/ m^2$  
&&   $\sim (M_{\rm Pl}/m)^3$ &\cr 
& Soliton Star:\cite{L87,LP87}  
&&  $10^{-2} (M_{\rm Pl}^4/ m \Phi_0^2)$  &&   
$2\times 10^{-3} (M_{\rm Pl}^5/ m^2 \Phi_0^3)$ &\cr 
height2pt&\omit&&\omit&&\omit&\cr}
\hrule}$$

In astrophysical terms, this maximal
mass is $M_{\rm crit} \cong 0.06\sqrt{\lambda}\, M^3_{\rm Pl}/m^2$ $=0.1
\sqrt{\lambda}$ ({\rm GeV/mc}$^2)^2\, M_\odot$. 
 
For light scalars, this value
can  even {\em exceed}  the limiting mass of 3.23 $M_\odot$
for a  neutron star (NS). 
For cosmologically relevant (invisible) axions of 
$m_{\rm a} \simeq 10^{-5}$ eV an axion star with the  rediculously
large mass of 
$M_{\rm crit} \sim 10^{27}\sqrt{\lambda}  M_\odot$ would be possible 
and stable \cite{SL97}.

For a {\em dilaton star} built from a very light dilaton $\chi$ of mass 
$m_{\rm dil}=10^{-11}$ eV/c$^2$,
Gradwohl and K\"albermann \cite{GK89} found 
\be
M_{\rm crit} = 7\sqrt{\overline\lambda} M_\odot\, , \qquad   
R_{\rm crit} = 40\sqrt{\overline\lambda}\; {\rm km}\,,
\ee
where $\overline\lambda$ is the rescaled coupling constant of the 
$\chi^4$ interaction.

Therefore, if scalar fields would exist in nature, such compact objects  
could even  question for massive compact objects 
the observational black hole paradigm in  
astrophysics.

\subsection{Stability and catastrophe theory}
For such soliton-type configurations kept together
by their self-generated gravitational field, the issue 
of stability is crucial. In the spherically
symmetric case, it was shown by Gleiser \cite{Ge88,GW89}, Jetzer 
\cite{Je90}, and Lee \& Pang \cite{LP89} that 
boson stars having masses below the Kaup limit 
are stable against small radial perturbation. 
More recently,
we have demonstrated via catastrophe theory \cite{kus,SKM92,KS92}
that this {\em stable branch} is even absolutely stable.
Moreover, our present surge stems 
from the possibility that these compact objects 
with a wide range of masses and radii could provide a
considerable fraction of the
non-baryonic part of dark matter \cite{SS94,sch}; see below.
Charged boson stars 
and their induced vacuum instabilities have been studied in 
Refs.~\cite{Je89,Je89b,JLS}. The problem of
non-radial pulsations of a boson star has  been mathematically
formulated \cite{KYF91}.

\subsection{Boson star formation}
The possible abundance of solitonic stars with astrophysical 
mass but microsco\-pic size could have  interesting implications 
for galaxy formation,  the
microwave back\-ground, and formation of protostars. 

Therefore it is an important question if boson 
stars can actually form from a primordial bosonic ``cloud" \cite{T91}. 
(The primordial formation of non-gravitating non-topological solutions was 
studied by Frieman et al.~\cite{FGGK}.)

\begin{figure}
\bec
\hskip0.01cm \psfig{figure=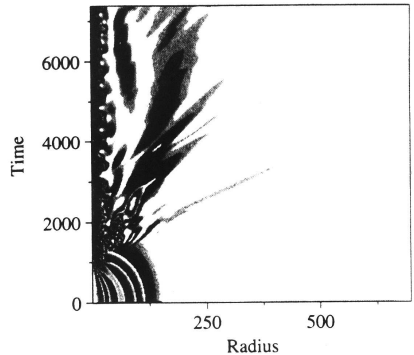,height=5cm}
\eec
\caption{Evolution of the shell density $r^2\rho$ for a 
massive, self--gravitating complex scalar field. 
Due to the self--generated gravity, the field collapses 
quickly and a perturbed boson star is formed.
\label{fig.2}}
\end{figure}

As 
Seidel and Suen \cite{SS90,SS94} have shown, cf.~Fig.~\ref{fig.2},
there exists a dissipationless 
relaxation process they call {\em gravitational cooling}.  Collisionless star 
systems are known to settle to a centrally denser system by sending some  
of their members  to larger radius. Likewise, a bosonic cloud will settle to a 
unique boson star by ejecting part of the scalar matter. Since there is no 
viscous term in the KG equation (\ref{phi153}), the radiation of 
the scalar field is the
only mechanism. This was demonstrated numerically 
by starting with a spherically symmetric configuration with 
$M_{\rm initial} \geq M_{\rm Kaup}$, i.e.~which is more massive then 
the Kaup limit. Actually such oscillating and pulsating branches have been 
predicted earlier in the stability analysis of Kusmartsev, Mielke, and 
Schunck \cite{KMS91,SKM92} by using 
catastrophe theory. Oscillating soliton stars were constructed by using
real scalar fields which are periodic in time \cite{SeSu91}.
Without spherical symmetry, i.e.~for 
$\Phi \sim R_a(r) Y_l{}^a(\theta\, ,\varphi) $, the emission of 
gravitational waves would also be necessary.

For a real (pseudo--) scalar field like the axion, the outcome is
quite different. 
The axion has the tendency to form compact objects 
(oscillatons) in a short time scale. Due to its intrinsic oscillations 
it would be, contrary to a BS, unstable. Since the field disperses to 
infinity, finite non-singular self--gravitating solitonic objects
cannot be formed with a massless Klein--Gordon field \cite{Ch86,P87},
but, instead, solutions with an infinite range can be found where the
mass increases linearly \cite{sch,Sch95,sch97}. These solutions can be
used to fit the observed rotation curves  for dwarf and spiral galaxies; see also
the contribution to the Dark Matter session of Schunck in these proceedings.
Similar investigations using excited BS states were used in
\cite{LK96,Sin95}.
In Ref.~\cite{KT93} a different mechanism for forming axion miniclusters 
and starlike configurations was proposed. For fermionic soliton stars, 
the temperature dependence in the forming of cold configurations 
has also been studied \cite{CV91}. 

\subsection{Gravitational waves} 

A boson star is an extremely dense object, since 
non-interacting scalar matter is very ``soft".  However, these   
properties are changed considerably by a 
{\em repulsive} 
self-interaction $U(\Phi)$. 

In the last stages of boson star  formation, one 
expects that first a highly excited configuration forms in which 
the  quantum numbers $n$, $l$ and $a$ of the gravitational atom, 
i.e.~the number $n-1$ of nodes, the 
angular momentum and the azimuthal angular dependence $e^{ia\varphi}$
are non-zero.  

In a simplified picture of BS formation, all initially high 
modes have eventually to decay 
into the ground state $n=l=a=0$  by a combined emission of scalar 
radiation and gravitational radiation.

In a Newtonian approximation \cite{FG89}, the energy released 
by scalar radiation from states with  zero quadrupole moment 
can be estimated by
\be E_{\rm rad} \sim (n-1) M_{\rm Pl}^2/m\; , \qquad 
\Delta N\sim (n-1)  (M_{\rm Pl}/m)^2\, .
\ee
This is accompanied by a loss of boson particles with the rate $\Delta N$ 
given above.

The lowest mode which has quadrupole moment and therefore 
can radiate {\em gravitational waves} is the $3d$ state 
with $n=3$ and $l=2$. For $\Delta j=2$ transitions,  it will decay into 
the $1s$ ground state with $n=1$ and $l=0$ while preserving the 
particle number $N$.   The radiated energy is 
quite large, i.e., $E_{\rm rad} =2.9 \times 10^{22}$ (GeV/mc$^2$) Ws.
Thus the final phase of the BS formation would terminate in an 
outburst of gravitational radiation despite the smallness of the object.

\subsection{Rotating boson stars} 
 
In recent papers \cite{miesch,Sch95,SM96,MS96}, we proved numerically that 
{\em rapidly rotating} boson stars with $a\neq 0$  
exist in general relativity. 
Because of the finite velocity of light and the infinite range of the
scalar matter within the boson star, our
{\em localized} configuration can {\em rotate only differentially}, but not
uniformly. Thus our new axisymmetric solution of the 
coupled Einstein--Klein--Gordon equations represent the 
{\em field-theoretical pendant} of 
rotating neutron stars which have been studied numerically 
for various equations of state and different approximation schemes 
\cite{fried,cook,[Er93]} as a model for (millisecond) pulsars; cf.~the paper
of Schunck and Mielke in these proceedings.

Kobayashi et al. \cite{koba} tried to find slowly rotating states ({\em near} the
spherically symmetric ones) of the boson star, but they failed.
The reason for that is a quantization of the relation of
angular momentum and particle number \cite{miesch}. In 
Newtonian theory, boson stars with axisymmetry have been  
constructed by several groups.
Static axisymmetric boson stars, in the Newtonian limit \cite{SB96}
and in general relativity (GR) \cite{YE97}, show that one can distinguish
two classes of boson stars by their parity transformation at 
the equator. In both approaches only the negative parity solutions
revealed axisymmetry, while those with positive parity merely  
converged  to
solutions with spherical symmetry.
The metric potentials and the components of the energy-momentum tensor
are equatorially symmetric despite of the antisymmetry of the scalar field.
In the Newtonian description,  Silveira and de Sousa \cite{SS95} 
followed the approach of Ferrell and Gleiser \cite{FG89} and constructed
solutions which have no equatorial symmetry at all. Hence, in GR, 
we have to separate solutions with and without equatorial symmetry.
In a more recent paper, Ryan \cite{R97} investigated the gravitational
radiation of macroscopic boson stars (with large self-interaction) by
 taking into account the reduction of the differential equations 
in this scenario.

\subsection{Gravitational evolution  and observation of boson stars}
Recently, several papers appeared which investigated the evolution
of boson stars if the external gravitational constant changes its value
with time \cite{T97,CS97,TLS97}; for an earlier investigation cf.~\cite{GJ93}.
This can be outlined within the theory of Jordan--Brans--Dicke
or a more general scalar tensor theory. The results show that
the mass of the boson star decreases due to a space-depending
gravitational constant, given through the Brans--Dicke scalar.
The mass of a boson star with constant central density is influenced
by a changing gravitational constant. Moreover, the possibility of a
gravitational memory of boson stars or a formation effect upon their
surrounding has been analyzed as well \cite{TLS97}.

The issue of observation has also been recently discussed \cite{SL97}.
Direct observation of boson
stars seems to be impossible in the near future.
But two effects could possibly  give indirect hints.
In the outer regions, the rotation velocity of
baryonic objects surrounding the boson star can reveal the star's mass.
Assuming that the scalar matter of the BS interacts only
gravitationally, we would have a transparent BS detecting a
gravitational redshift up to values of $z=0.68$ observable by
radiating matter moving
in the strong gravitational potential. For further investigations of
rotation curves, cf.~Ref.~\cite{Sin95,LK96}.

\section{Dilaton stars and kinks}  
Real massless scalar fields coupled to Einstein gravity 
are known to admit {\em exact} solutions. 
Already in 1959, Buchdahl \cite{B59}
found a continous two--parameter family of static solutions:
Accordingly, any static vacuum solution $g_{\mu\nu}=
(g_{00}\, , g_{AB})$ 
of Einstein's theory can be mapped into the solution 
\be
\cases{
(g_{00}^\beta \> , g_{00}^{1-\beta} g_{AB})\cr 
  \Phi =\lambda \ln g_{00} }
  \quad {\rm with}\quad  \beta=\pm \sqrt{1-2\lambda^2} 
\ee
 of the 
Einstein--KG system. For an  extension to 
 conformally coupled scalar fields, see Bekenstein \cite{B74}.

In the framework of the Jordan--Brans--Dicke--Thiry theory, these solutions  
appear already  in Ref.~\cite{E55} and correspond to those found by 
Majumdar \cite{M47} for the Einstein--Maxwell system. Later they were 
rederived by Wyman \cite{W81} and for a special case recovered 
independently by Baekler et al.~\cite{BMHH87}.
Further closed analytical expressions of the 
so-called Wyman solution \cite{W81} are constructed via Computer Algebra 
in Ref.~\cite{SS91}. Generalizations to spacetimes of arbitrary dimensions 
are reconsidered by Xanthopoulos and Zannias \cite{XZ89} in the 
spherically symmetric isotropic case, see also Ref.~\cite{XD92}
in the case of a conformally coupled scalar field.
The global initial value problem for a self--gravitating massless real scalar 
field has been analyzed by Christodoulou \cite{Ch86} in a 
spherically symmetric spacetime, see also Choptuik \cite{Ch93}. 
G\"urses \cite{G77} found conformally flat solutions. For all solutions the 
scalar field develops a logarithmic singularity at the origin,
for some solutions the metric 
becomes there also singular leading to a {\em naked singularity}.

A generally relativistic Klein--Gordon field with an effective  
$\Phi^3$    self-interaction for an interior ball has also been 
analyzed  \cite{CKT77}. In order
to avoid a singular configuration at the origin, a repulsive 
(or ``ghostlike") scalar field has been chosen as a source of Einstein's 
equations. In a further
step,  Kodama et al.  \cite{KCT78,K78,KOS79} constructed
spherically symmetric {\em kink-type} 
solutions for a 
repulsive scalar field with a $U(\Phi)\sim (const. -\Phi^2)^2$ 
self-coupling (compare also with 
Ellis \cite{E79}).
As is common for kinks, the radial function at spatial infinity is chosen
to be $\pm$ 
const.~characterizing this nonlinear model. 
The  constant is necessary in order to eliminate the induced 
cosmological constant
which otherwise would occur for the constant solution characterizing
the kink solution asymptotically. 

These type of solution, however, suffer from a dynamical instability, see 
Jetzer and Scialom \cite{JS92}. 
In flat spacetime, according to Derrick's theorem \cite{De64},
no stable time-dependent solutions 
of finite energy exist for a non-linearly coupled real scalar field.
 
Other scalar fields arise from axion \cite{KT93}, inflaton or dilaton
fields \cite{GK89} with there corresponding compact objects.
Recently, stationary axisymmetric solution of the 
Einstein--dilaton--axion action 
are obtained by Garc\'{\i}a et al.~\cite{GGK95}.

In the process of a Kaluza--Klein type dimensional reduction of supergravity 
or superstring models there arises the dilaton field $\chi$ as part of the 
higher--dimensional 
metric. These real scalar fields couple to gravity in the non--minimal 
$\chi^2 R$ fashion, resembling the Brans--Dicke field of scalar--tensor theories. The 
corresponding {\em dilaton stars} \cite{GK89} are  stable because of a {\em conserved 
dilaton current} and charge
$Q_{\rm dil}$ in such models.

In an extended model  \cite{TX92} with Higgs field $\Phi^I$ and 
dilaton coupling, there 
occur, however, unstable branches with diverging mass $M$ for high 
central values $\vert\Phi^I(0)\vert$ of the 
Higgs field.

\section{Other gravitational solitons}

To some extent Wheeler's concept of  geons \cite{W55}  has anticipated the 
(nonintegrable) 
solition solutions \cite{M78} of classical nonlinear field theories. As 
mentioned in the Introduction, a geon or gravitational soliton 
originally was meant to consist
of a spherical shell of electromagnetic radiation held together by its 
own gravitational attraction. In the idealized case of a thin spherical 
geon, cf.~Pfister \cite{P91}, the corresponding metric functions have the 
values $\exp(\nu_c) =1/9$ well inside 
and $\exp(\nu) =  \exp(\lambda) =1- 2m(r)/r$           
well outside the active region. The trapping area for the electromagnetic wave
trains has a radius of $r_{\rm active}= 9m/4$. This result has been confirmed by applying
Ritz variational principles \cite{E57}.
Although this procedure is rather artifical, thereby one obtains  
a ``bag--like" object \cite{CJ74} having inside a portion of an 
Einstein microcosmos and outside a 
Schwarzschild manifold as background spacetime.

Configurations with toroidal or linear electromagnetic waves have 
been constructed 
by Ernst \cite{E57}, the cylindrical geons of Melvin \cite{M64} are stable 
against gravitational collapse under large radial perturbations.  
Neutrino geons have been analyzed by 
Brill and Wheeler \cite{BW57}. 
Brill and Hartle \cite{BH64} could even demonstrate the existence of 
gravitational 
solitons constructed purely from gravitational waves. By expanding 
the occurring gravitational waves in terms of tensor
spherical harmonics,  it can be  shown \cite{RW57} that the radial 
function experiences the
same effective potential  except that an additional factor   appears in 
front of the contributions from the background metric.
In a recent paper,
the possibility of {\em black holes} formed by collapsed gravitational waves
has been discussed \cite{holz,whee95}. Although these 
objects are weakly unstable, they could contribute to dark 
matter \cite{holz}.

For the generally relativistic kink of Kodama \cite{K78}, 
the radial solution  becomes zero
at a certain radius $r_0$   at which the background geometry develops a 
Schwarzschild type horizon. (Geon--type solutions exhibiting an 
event horizon may be termed ``black solitons" \cite{SS76}.) The boundary 
condition at $r_0$,  however, allows an 
extension of these solutions into a three-manifold consisting of two 
asymptotically Euclidean spaces connected by an 
Einstein--Rosen bridge \cite{ER35}. Arguments 
are given that this extended, nonsingular configuration is stable 
with respect to radial oscillations.
It should be noted that such solutions cannot be constructed for the 
wormhole \cite{M77a}
topology $R\times S^1\times S^2$   which would be obtainable by 
identifying the asymptotically flat regions. The reason simply being that 
the radial functions of the kink has an opposite sign in the other sheet 
of the Universe.

New wormhole type solutions are discussed by Ellis \cite{E79} 
and by Thorne et
al.~\cite{MTY88}, cf. also Ref.~\cite{ABT}.
Their throats  will, however, be kept open by ``exotic matter" 
which violates the 
weak energy condition $T_{\mu\nu}u^\mu u^\nu \geq 0$  for 
timelike  vectors $u^\mu$. Then such wormhole configurations  
would allow closed timelike curves and the 
paradoxical possibilty of a ``time machine" \cite{MTY88}.

Although we have no intention to give a complete review, we would like to 
mention that other studies on geons involve massless scalar fields 
\cite{DH70,B74,G77,LWT80}, coupled
Einstein--Maxwell--Klein--Gordon systems \cite{D63,BM79,TWS75,BD77}, 
the generally relativistic Dirac equation in an external gravitational field
\cite{Mi81,Vi90}, or 
even combined Dirac--Einstein--Maxwell
field equations \cite{HD77}.

According to a result of 
Brill \cite{B64}, a massless scalar field can even be geometrized in 
the sense of the already unified field
theory or geometrodynamics of Rainich, Misner, and Wheeler \cite{W62}. 
Loosely speaking,
this means that the scalar field can be completely read off from 
the ``footprints" it leaves on the geometry.

The non-topological solitons (NS) of Rosen \cite{Ro68} 
as well as of Lee and Wick \cite{LW74,FLS} can be regarded as 
the non-gravitational precursors
of boson stars. For a specific Higgs type 
self-interaction $U(\Phi)$, they are localized solutions of a 
non-linear Klein-Gordon equation in flat spacetime. Spherically symmetric 
solutions in a prescribed gravitational background such as 
that of Schwarzschild or constant curvature were presented 
in Refs.~\cite{El88,DM79,O78,Mi80}.

Similar configurations are called  Q--balls \cite{Co85}, 
which are stabilized by 
the conserved (baryon number) charge $Q$, fermion Q--balls \cite{BLS90},
 neutrino balls \cite{Ho87}, and quark nuggets \cite{Wi84} 
in the case of spinors. Bound further by their self-generated
gravitational field, such Q--stars may model 
neutron stars  with an equation of state usually not accessible  in the 
laboratory. Therefore, their mass  can also exceed the 
Chandrasekhar limit of $\sim 3\, M_\odot$ for neutron stars.

In quantum chromodynamics (QCD), nowadays the most prominent model for strong 
interactions, the dynamics of the mediating vector gluons is determined 
by an action modelled after Maxwell's theory of electromagnetism. The 
resulting model is a gauge theory of the Yang--Mills type. 
However, it is known \cite{CS77} that in such 
sourceless non--Abelian gauge theories there are no {\em classical 
glueballs}  \cite{R77}
which otherwise would be an indication for the occurrence of confinement in 
the quantized theory. 
The reason 
simply is that nearby small portions of the Yang--Mills fields 
always point in the same direction in internal space and therefore must 
repel each other as like charges. 

Monopole type solution of the Einstein--Yang--Mills system are 
found numerically by 
Bartnik and McKinnon \cite{BM87} in which gravity balances the 
repulsion of the internal gauge fields. An interesting attempt to 
determine the solution analytically 
in terms of a series expansion and nonlinear recursion relations is 
given by Schunck \cite{Sch93}.

\section*{Acknowledgments}
We would like to thank John D.~Barrow, Andrew R.~Liddle, and 
Alfredo Mac\'{\i}as
for useful discussions,  literature hints, and support.
This work was partially supported by  CONACyT, grants No. 3544--E9311, No. 
3898P--E9608, and by the joint German--Mexican project 
DLR--Conacyt 6.B0a.6A. 
One of us (E.W.M.) acknowledges the support by the short--term
fellowship 961 616 015 6 of the German Academic Exchange Service (DAAD), Bonn.
F.E.S.~was supported by an European Union Marie Curie TMR fellowship.

\nonfrenchspacing

\end{document}